\documentclass[aip, jcp, showpacs, superscriptaddress, preprint ,11pt]{revtex4-1}
\usepackage{dcolumn,amsmath}
\usepackage{longtable}
\usepackage{graphicx}
\usepackage{multirow} 
\usepackage{amsmath}
\usepackage{bm}
\usepackage{comment}
\usepackage[export]{adjustbox}
\newcommand{\wn}{cm$^{-1}$}
\newcommand{\Np}{$J^{\prime}$}
\newcommand{\Kp}{$K^{\prime}$}
\newcommand{\Ep}{$E^{\prime}$}
\newcommand{\Npp}{$J^{\prime \prime}$}
\newcommand{\Kpp}{$K^{\prime \prime}$}
\newcommand{\Epp}{$E^{\prime \prime}$}
\newcommand{\supa}{$^a$}
\newcommand{\supb}{$^b$}
\newcommand{\supc}{$^c$}
\newcommand{\supd}{$^d$}
\begin{document}

\title{Quadrupole splittings in the near-infrared spectrum of $^{14}$NH$_3$}

\author{Sylvestre\  Twagirayezu}\email{stwagirayezu@lamar.edu}\thanks{Present Address: Department of Chemistry \& Biochemistry, Lamar University, Beaumont, TX 77710}
\author{Gregory\ E.\ Hall}\email{gehall@bnl.gov}
\affiliation{Division of Chemistry, Department of Energy and Photon Sciences, Brookhaven National Laboratory, Upton, NY 11973-5000, USA}
\author{Trevor\ J.\ Sears}\email{sears@bnl.gov, trevor.sears@stonybrook.edu}
\affiliation{Division of Chemistry, Department of Energy and Photon Sciences, Brookhaven National Laboratory, Upton, NY 11973-5000, USA}
\affiliation{Chemistry Department, Stony Brook University, Stony Brook, NY 11794, USA}

\begin{abstract}

Sub-Doppler, saturation dip, spectra of lines in the $v_1 + v_3$, $v_1 + 2v_4$ and $v_3 + 2v_4$ bands of $^{14}$NH$_3$ have been measured by frequency comb-referenced diode laser absorption spectroscopy.  The observed spectral line widths are dominated by transit time broadening, and show resolved or partially-resolved hyperfine splittings that are primarily determined by the $^{14}$N quadrupole coupling.  Modeling of the observed line shapes based on the known hyperfine level structure of the ground state of the molecule shows that, in nearly all cases,  the excited state level has hyperfine splittings similar to the same rotational level in the ground state.  The data provide accurate frequencies for the line positions and easily separate lines overlapped in Doppler-limited spectra.  The observed hyperfine splittings can be used to make and confirm rotational assignments and ground state combination differences obtained from the measured frequencies are comparable in accuracy to those obtained from conventional microwave spectroscopy.  Several of the measured transitions do not show the quadrupole hyperfine splittings expected based on their existing rotational assignments. Either the assignments are incorrect or the upper levels involved are perturbed in a way that affects the nuclear hyperfine structure.         

\end{abstract}

\date{\today}

\maketitle

\section{Introduction}
 Ammonia, NH$_3$, is  a symmetric top rotor possessing a large amplitude inversion vibrational mode.  The inversion motion is associated with a double-well potential\cite{Kroto} with  an effective barrier height of 2020 \wn.  Tunneling though the barrier leads to the well-known inversion splittings\cite{Kroto, Townes} in the zero point, and higher, levels of the molecule.  Further, the spins of the three equivalent protons result in two nuclear spin states of the molecule, ortho- ($\bm{I_H} = 3/2$) and para- ($\bm{I_H} = 1/2$) each associated with a distinct set of rotational levels.\cite{Bunkerbook} The two sets of levels do not interact under most circumstances.  These characteristics, together with its practical importance, have made ammonia a prototype molecule for spectroscopic studies. \\ 
 
 There has been considerable published work on the near-infrared spectrum of ammonia during the past few years.\cite{Sung2012, Cacciani2012, Foldesi2014, AlDerzi2015, Lehmann, Douglas, Czajkowski2009, Berden1999} The spectrum in this region consists of a mixture of combination and overtone bands, of which the perpendicular bands $v_1 + v_3$ , $v_1 + 2v_4$, $v_3 + 2v_4$,  and  $2v_3$ are the most prominent.  However, despite many years of work, spectroscopic analysis remains incomplete  because of the many overlapping rotational features even at Doppler-limited resolution, multiple uncharacterized perturbations caused by anharmonic and Coriolis mixing, and the presence of hot band lines in the low-frequency inversion mode.  The great majority of the observed features in recent high resolution measurements in the near-infrared remain rotationally unassigned.\cite{Sung2012, Cacciani2012}  Isotopically labeled samples\cite{Lees2008} permitted assignments of some of the strongest features of $^{14}$NH$_3$ and $^{15}$NH$_3$ spectra, but again leave many features unassigned.  Variable (low) temperature studies\cite{Foldesi2014} of the spectrum have led to estimates of the lower state energies for many of the spectroscopic features and have permitted further assignments and corrections to earlier ones. Very recently, the Tennyson group\cite{AlDerzi2015} critically reviewed and validated all the available high resolution spectroscopic data for ammonia, to create a database of known levels and band origins.  Combined with previously calculated absorption intensities, this work provides a detailed and accurate map of the assigned spectrum of ammonia from the microwave though the near-infrared.  The work resulted in a Measured Active Rotational-Vibrational Energy Levels (MARVEL) database that can be dynamically updated as new data become available.  \\

In the present work, we report sub-Doppler measurements of a number of transitions in the $v_1+v_3$, $v_1+2v_4$ and $v_3+2v_4$ bands.  These measurements fully resolve many overlapped vibration-rotation features in the Doppler-limited spectra whose presence have previously hindered assignments.\cite{Lees2008} The new measurements also exhibit partially resolved nuclear hyperfine structure due to the $^{14}$N quadrupole and proton hyperfine couplings. The observed hyperfine patterns are spectroscopic signatures of the rotational levels involved in the transitions\cite{Kukolich67, Kukolich68, Kukolich70, Hougen72} and may therefore be used to confirm or deny spectroscopic assignments. Previous saturation spectroscopy measurements of ammonia at these wavelengths, carried out on a sample in a hollow-core fiber, have been reported.\cite{Cubillas2008, Petersen2010}  The resolution reported was sufficient to resolve lines blended in the Doppler-limited spectrum, but not sufficient to resolve any hyperfine structure.   Czajkowski et al.\cite{Czajkowski2009} have also previously reported the measurement of a few sub-Doppler lines in the ammonia spectrum in this region, but again did not report the observation of quadrupole structure, presumably because the measurements were conducted at higher pressures than the current work, resulting in some collisional broadening. \\

Two of the  transitions observed, the $v_1+v_3$ band, $^pP(J,K=5,4)_a$ line at 6537.6806 \wn, frequency measured here at 195 994.734 57(2) GHz, and the $(v_1+2v_4)$, $^RP(7,5)_a$ line at 6488.200 \wn, 194 511.328 21(2) GHz here, exhibit hyperfine structure that does not conform to that expected, based on the assumption that the known lower state hyperfine splittings\cite{Hougen72} do not change on vibrational excitation.  Sub-Doppler measurements of other transitions to confirm the relevant ground state combination differences are needed.  For now, these assignments, based on Doppler-resolved spectra,\cite{Sung2012, AlDerzi2015} must be regarded with suspicion.  If the rotational assignments are correct, the observed change in the quadrupole splitting must be due to a perturbation in the upper level involved in the transitions.  Examination of the upper energy levels listed in the MARVEL database,\cite{AlDerzi2015} shows that the $v_1+v_3$, $(4,3)_a$  level is within approximately 0.04 \wn  in energy of $(5,4)_s$ of  $(v_1+2v_4)$.  The level $(4,3)_a$ of $(v_1+v_3)$ belongs to the para- set of proton hyperfine levels,\cite{Bunkerbook} while  $(5,4)_s$ of $(v_1+2v_4)$ is an ortho- level.  However, the estimated magnitude of the ortho-para coupling terms in ammonia\cite{Cacciani2009} is too small to account for such a (relatively) large observed perturbation between these levels.  In future studies of the spectrum of ammonia in this region, the possibility of ortho-para coupling between nearly degenerate levels of the appropriate symmetry should be kept in mind, but accidental degeneracies close enough to allow significant ortho-para mixing are likely to be rare.

\section{Experimental Methods}
   The spectrometer used in this work has been described in detail previously.\cite{Twagirayezu1} Samples of anhydrous ammonia gas (Matheson Gas, Inc.) were introduced in the cavity-type absorption cell at pressures of between 1 and 20 mTorr (0.133-2.66 Pa) depending on the strength of the absorption. Sub-Doppler, saturation dip, spectra of rotational lines in the ammonia spectrum near 1.5$\mu$m were recorded.  Collisional broadening of the saturation features could be observed at higher pressures, but no attempt was made to quantify this effect in this work. As in previous work,\cite{Twagirayezu1} the stronger saturation dip features could also be power-broadened and distorted at higher laser powers. Therefore, spectra were recorded at the lowest pressure and laser power consistent with satisfactory signal-to-noise ratios. Examples of observed derivative signals of the saturation dip profiles are shown in Figures \ref{fig1} and \ref{fig2}. These were obtained by scanning  the comb repetition rate by 0.25 Hz/step, corresponding to approximately 60 kHz/step in the optical frequency across the saturation dip, typically collecting an averaged signal for 3 seconds at each frequency step. A few weak lines were recorded with more averaging to obtain satisfactory signal-to-noise ratios. Note that the modulation depth used to record Figure \ref{fig2} was reduced compared to that shown in Figure \ref{fig1} to reduce modulation broadening and highlight the splittings.  The figures also show the results of the line analysis described in detail below.  
 
\section{Results and Analysis}
  \subsection{Hyperfine Level Structure}
     The observed saturation line shapes arise from partially resolved hyperfine splittings.  The dominant hyperfine splitting is due to the nuclear quadrupole of the $^{14}$N nucleus, $\bm{I_N}=\bm{1}$,  splitting each rotational level ($J>0$) into three sub-levels, labeled here by the quantum number $F_1$ with $F_1$\, =\, $J$, $J\pm1$.  Superimposed  
      on the quadrupole structure, each level is further split by the proton nuclear hyperfine structure due to the $^1$H nuclear spin-rotation coupling and nuclear spin-spin dipolar coupling. The three equivalent protons lead to ortho- ($\bm{I_H}=\bm{3/2}$) and para- ($\bm{I_H}=\bm{1/2}$) proton nuclear spin functions,\cite{Bunkerbook} and the total angular momentum is $\bm{F}\, =\, \bm{F_1}+\bm{I_H}$, quantum number $F$.  Measurements by Kukolich\cite{Kukolich67, Kukolich68, Kukolich70} and additional analysis by Hougen\cite{Hougen72} fully characterized the hyperfine structure in the lowest inversion doublet ($v_2 = 0$) of the $^{14}$NH$_3$ molecule.  However, the proton hyperfine splittings are less than a few tens of kHz, too small to be resolved in the current experiments.   The quadrupole coupling energy contributions are determined by the matrix elements:\cite{Kukolich67, Hougen72}
\small
\begin{equation}\label{WQ1}
     W_Q(F_1,J^{\prime}, J, K) = (-1)^{J+I_N+F_1}  { \begin{Bmatrix}  F_1 & I_N & J' \\ 2 & J& I_N \end{Bmatrix} }
       <J'~K\mid V \mid J~ K> < I_N\mid\mid Q \mid\mid I_N >
 \end{equation} 
 \normalsize
 where V and Q  are matrix elements which depend on $J^{\prime}$, $J$, $K$ and $I_N$.  The $V$ contribution is: 
 \small
 \begin{equation}\label{V1}
      <J'~K\mid V \mid J~ K>= \frac{1}{2} q\left[(2J'+1)(2J+1) \right]^{1/2}  
                   (-1)^{J'+ K} \begin{pmatrix} J' & 2 & J \\ -K & 0 & K \end{pmatrix}      
\end{equation}
\normalsize
  and $Q$:
\small 
\begin{equation}\label{Q1} 
     < I_N\mid\mid Q \mid\mid I_N > = \frac{1}{2} eQ (2I_N+1)\left[ \frac{(2I_N+3)(I_N+1) }{I_N(2I_N-1)(2I_N+1)} \right]^{1/2}     
\end{equation}\\
\normalsize
   The energy contributions from matrix elements off-diagonal in $J$ in eq. (\ref{WQ1}) are negligible compared to the experimental resolution here because they connect levels separated by a rotational term value.  Therefore, we computed just the diagonal contributions to the quadrupolar splittings in simulating the observed spectral features \textit{i.e.}  $J'$=$ J$ in eq. (\ref{WQ1}) and (\ref{V1}) above. These expressions were used to calculate the quadrupolar hyperfine splittings for each rotational level of interest.  Also, although the inversion doublet levels have slightly different quadrupolar splittings, determined by the parameter $\Delta Q^*$ in Hougen's notation, \cite{Hougen72} these differences are small compared to the current experimental measurement precision, so they too were neglected. Hence, all the experimental splittings will depend on the one quadrupolar parameter $\frac{eQq}{4}$, assumed to be the same in both the ground and excited vibrational levels. The calculated $^{14}$N quadrupole splittings based on this model for the rotational levels in NH$_3$ of interest here are given in Table \ref{table1}. We note that the splittings for the $(J,K) = (3,2)$ levels are zero in this approximation because of the factor $J(J+1)-3K^2$ arising from the expansion of the 3-$j$ symbol in equation \ref{V1}. \\
   
 \subsection{Saturation line profiles}
   Spectroscopic transitions are expected to obey the selection rules, $\Delta F_1= 0,\pm1$ and the observed line shapes were modeled as a convolution of Lorentzian derivatives with an adjustable modulation broadening\cite{Axner2001} for each possible quadrupolar hyperfine transition as detailed below.  Relative intensities of the quadrupole split transitions are given by:
   \small
  \begin{equation}\label{IntQ}
    Intensity_Q(rel.) =\ (2F'_1+1)(2 F_1+1) \begin{Bmatrix}  I_N & J'& F'_1 \\ 1 & F_1& J \end{Bmatrix} ^{2}
  \end{equation} 
  \normalsize
 
   Here, $J^{\prime}$ and $F^{\prime}$ are the quantum numbers in the upper state of the transition.  With the aid of eq.(\ref{WQ1}) and eq.(\ref{IntQ}), the observed line shapes were modeled in steps: (i) the quadrupole  energy level patterns for the rotational levels involved in the transition were estimated as described above (ii) the component quadrupole transition frequencies were then obtained  by subtracting the lower energy levels ($W_Q(F_1,J,K)$) from the upper split levels ($W_Q(F'_1,J',K')$) with the selection rules, $\Delta F_1 = 0,\pm1$, and the relative intensities computed from eq. (\ref{IntQ}); (iii) crossover resonances, due to pairs of two-level transitions sharing a common upper or lower hyperfine component, were also computed with a resonance halfway between the contributing two-level saturation frequencies and an intensity given by the geometrical mean of their intensities.\cite{Demtroder}  Finally, (iv) the calculated saturation resonance frequencies and relative intensities were used to create an overall simulated line shape, from an intensity weighted sum of Lorentzian derivative lines using the Axner \textit{et al.} model,\cite{Axner2001} each with a width determined by the estimated transit-time broadening (HWHM = 290 kHz) and an empirical modulation broadening to account for other broadening as detailed below, to compare to the experimental trace. \\
   
\begin{figure*} [h!]  
  \centering
  \includegraphics[width=0.4\textwidth]{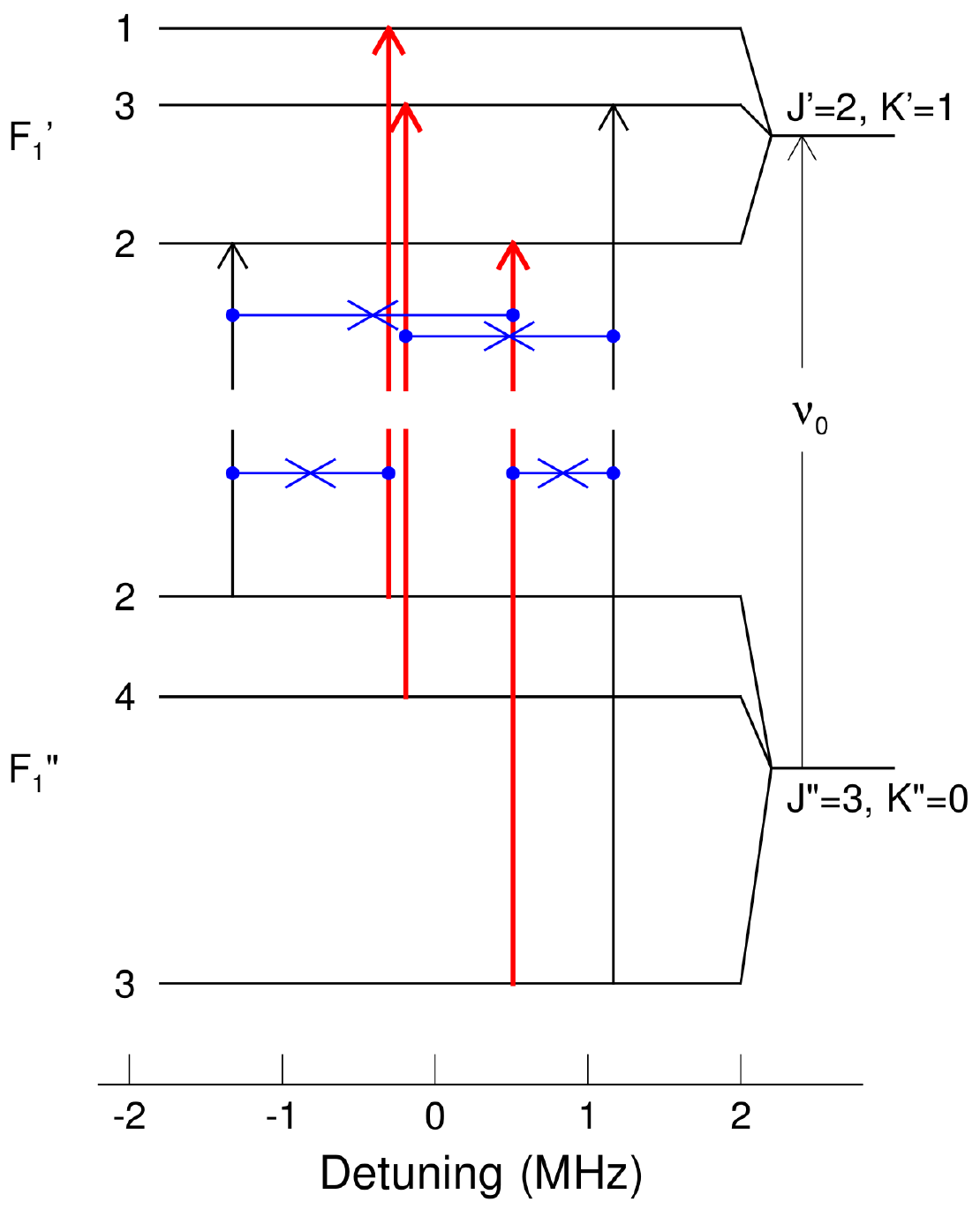}%
  \centering
  \includegraphics[width=0.6\textwidth]{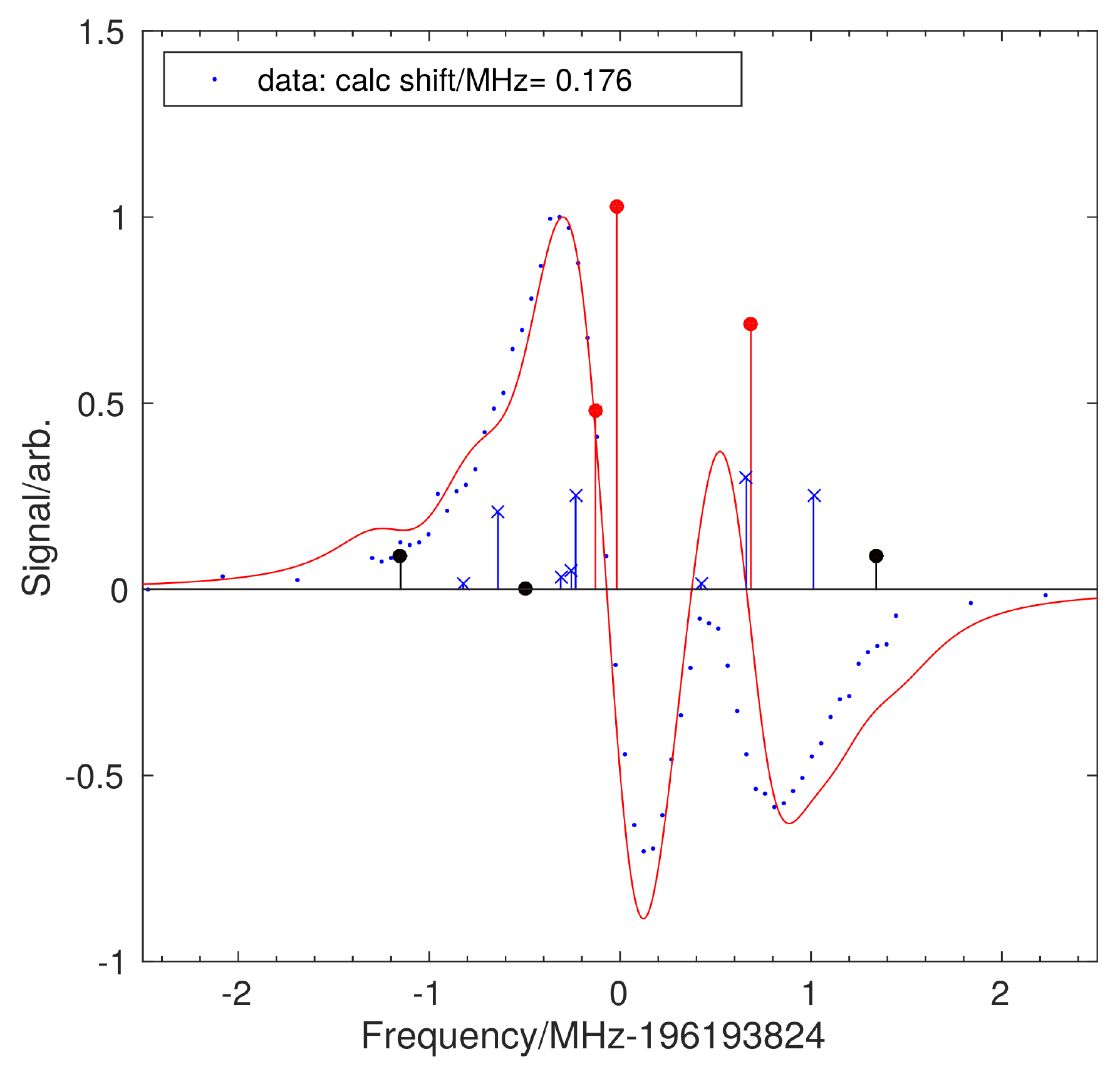}%
\caption{\footnotesize{Quadrupole hyperfine level structure for rotational states involved in the $^rP(3,0)_s$ transition of the $v_1+v_3$ band at 196 193.824 176(30) GHz.  Left panel shows $^{14}$N quadrupole splittings and the strong (red), weaker (black) and crossover (blue $\times$) transitions, plotted at the corresponding detunings from the hyperfine-free transition frequency, $\nu_0$. The transition from $F_1''$ = 2 to $F_1'$ = 3 is not forbidden, but has negligible intensity, and is not shown in the left diagram. The right panel shows the saturation dip line shape (points) observed at a pressure of 5 mTorr and a one-way intra-cavity power of 100 mW.  The red, black and blue stick spectra correspond to the same color scheme, with heights proportional to the calculated intensities.  The transition frequency $\nu_0$ is adjusted to optimize the agreement of observed and simulated line profiles at the dominant zero crossing.}}%
\label{fig1}
\end{figure*}

\begin{figure} [h] 
  \includegraphics[width=0.5\textwidth] {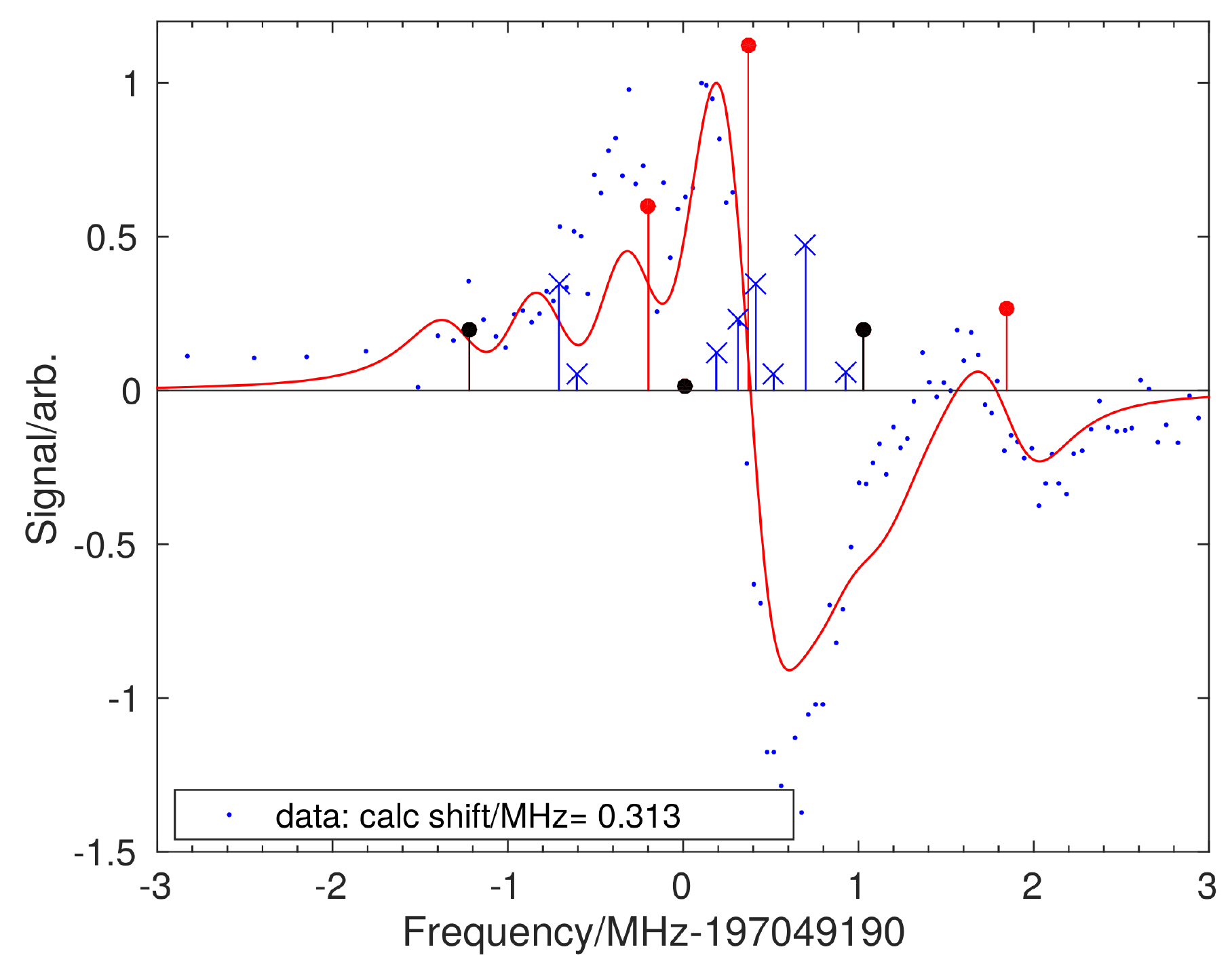} 
  \centering
 \caption{\footnotesize{$^pP(2,1)_s$ transition of the $v_1+v_3$ band at 197 049.190 313(30) GHz. The dots are experimental data, and the solid red line is the modeled line shape. The stick spectra represent the center positions and relative intensities of the components contributing to the line shape.  As in Fig. 1, the main  $\Delta F_1 = \Delta J$ lines are shown in red,  $\Delta F_1 \neq \Delta J$ lines are shown in black, and crossover transitions are indicated with a blue line and $\times$ symbol. }}
 \label{fig2}
 \end{figure}

   Figure \ref{fig3} illustrates the variety of quadrupolar split hyperfine patterns that can be expected for $P$, $Q$, and $R$ type transitions with $\Delta K=0, \pm 1$, given the experimental resolution of this work. For purposes of illustration, the line shapes for transitions originating in the state  $(J,K)=(5,3)$ were computed, with only the three dominant  $\Delta F_1 = \Delta J$ hyperfine transitions included for each rotational line. Ammonia only has perpendicular bands in this spectral region of interest, so the examples in the center column of the figure are not expected to be observed.  Experimentally, we have found transitions involving rotational levels with $J > 3$ are reasonably well modeled by the line shapes in this figure.  For transitions involving lower rotational levels, additional hyperfine components, \textit{i.e.} $\Delta F_1 \ne \Delta J$,  and crossover resonances need to be included to model the observed shapes more reliably.  Examples are shown in Figs. \ref{fig1} and \ref{fig2}.  \\   
   
    All the measured  saturation  dip transitions showed either resolved splittings or distortions compared to a wavelength-modulated derivative of a single Lorentzian transmission line shape.\cite{Axner2001}   Individual line widths are dominated by transit-time broadening,\cite{Twagirayezu1} but poorly defined contributions from power, collisional (at pressures $>$5 mT) and modulation broadening also contribute, and we found that the chosen function provided a better approximation to the observations compared to the expected Gaussian resulting from purely transit-time broadening effects.\cite{Demtroder} \\
   
   The calculated profile was then matched to the observed feature with an adjustable central frequency ($\nu_0$) and the fixed set of quadrupolar offsets, $\Delta$W$_Q$, derived from the line assignment and the quadrupole splittings in table \ref{table1}. Except in a few cases, discussed below, the predicted line shapes were in good accord with the observed features.  The predictions might be refined by slightly adjusting the estimated quadrupolar shifts ($\Delta$W$_Q$) for the component transitions by varying the upper, quadrupole split, energy levels while keeping ground state splittings fixed.  However, adjusting the upper state quadrupole coupling parameter to fit one or two poorly modeled line shapes would result in worse agreement with the great majority of the measurements.  Figure \ref{fig1} illustrates the simulation results for the $^rP(3,0)_s$ transition at 196 193.824 GHz\cite{Sung2012, AlDerzi2015} while Figure \ref{fig2} illustrates the more complex line shape observed for a $^pP(2,1)_s$ transition.  \\ 
   
   In Fig. \ref{fig1}, the calculated hyperfine pattern is close to that observed, but the match would be improved if the $F_1^{\prime}=2$ component is slightly lower in energy.  For the $^pP(2,1)_s$ transition in Fig. \ref{fig2}, the the positions of the hyperfine components lie close to the observed positions, but the relative intensities of the weaker components are underestimated, or the strongest component overestimated, compared to the observed.  We attribute this is varying degrees of saturation for the different components, but trials using a model explicitly varying the relative intensities based on their linear line strengths were not successful in improving the qualitative agreement. Figure \ref{fig4a} shows examples of a series of transitions from levels with $J=5$.  The difference between the $^pP(5,4)$ $a-$ and $s-$ components is noteworthy.  In the model, the two should have identical shapes, and the observations for the $K=5$ and $K=3$ transitions match this expectation. \\
   
   \subsection{Hyperfine-free rest frequencies}
   Values of the hyperfine-free rest frequencies $\nu _{0}$ were extracted for each frequency measured line and summarized in Table \ref{table2}.  The offset between $\nu _{0}$ and the dominant zero-crossing frequency of the partially resolved hyperfine pattern comes from the line shape modeling, so that even though the observed central zero crossing frequency is determined with high precision, the reported $\nu _{0}$ value will include an additional model-dependent error. These errors are of the same order of magnitude as errors expected from the neglect of the proton hyperfine and other neglected contributions to the line shape and are close to the experimental resolution.  For this reason, we report a conservative estimate of the line center measurement errors as $\pm 20$ kHz for most of the data.  Exceptions are noted in table \ref{table2} when the lines were weak, for transitions where the observed quadrupole splitting patterns were complicated, typically low-$J$ transitions, and for those transitions where the predicted line shape did not match the observed.  The accuracy of the measurement of the position of the zero-crossing of an observed line shape is actually better than this, and of the order of 3-10 kHz.\cite{Twagirayezu1} \\

\begin{figure} [h] 
  \includegraphics[width=0.5\textwidth] {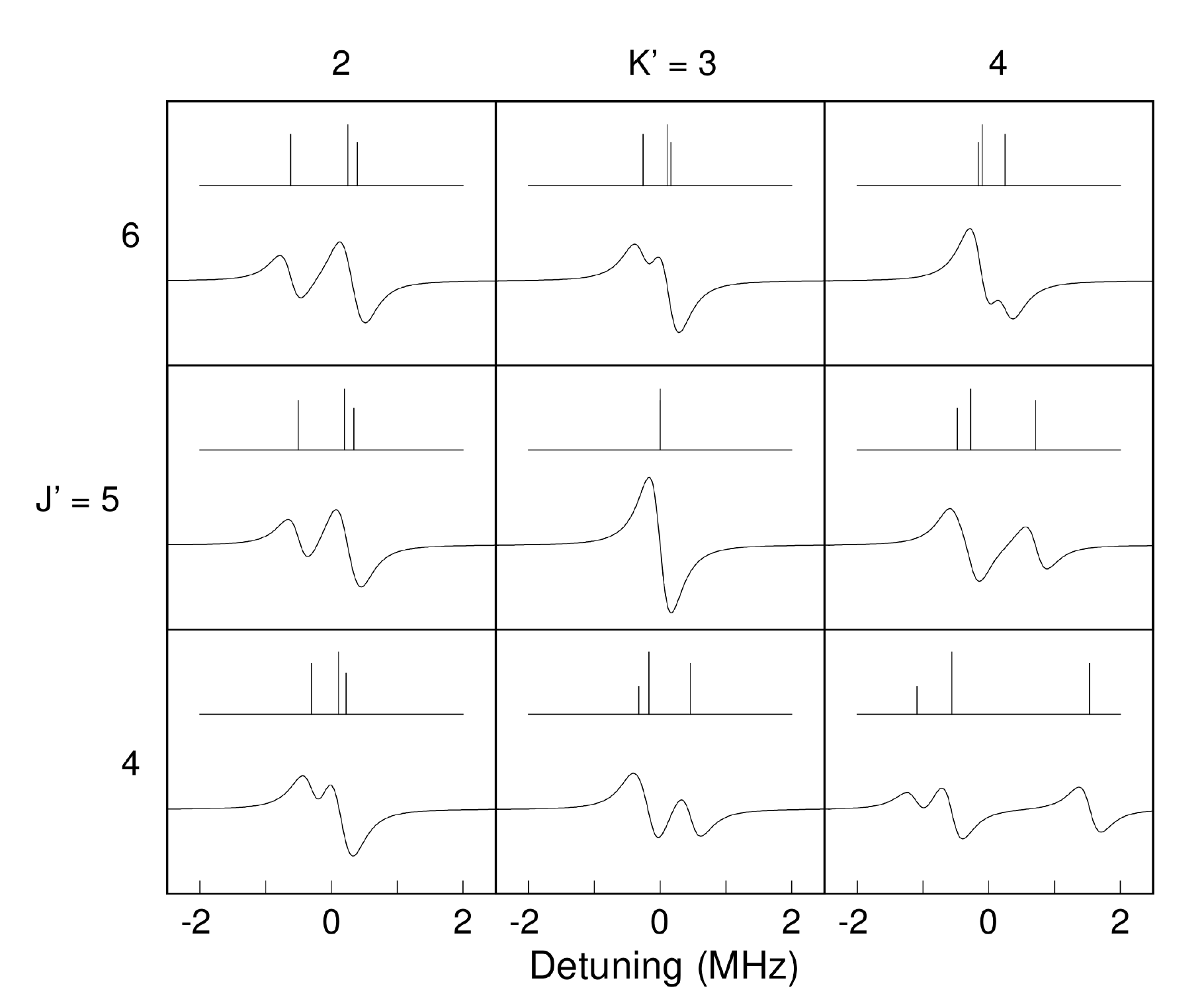} 
  \centering
 \caption{\footnotesize{Illustration of the variety of quadrupole splittings expected for different rotational transitions coming from a single rotational level.  $(J'',K'')=(5,3)$ is used as an illustration.  The modeling assumes a derivative Lorentzian broadened line shape function with width (HWHM) of 290 kHz, appropriate to that observed.  For these simulations, only the strongest components, \textit{i.e.} $\Delta F_1 = \Delta J$ are included.  Crossover resonances are not included; they have only a small effect for these rotational levels.  The center column  is included for completeness; these parallel transitions are not present in the current data. }}
 \label{fig3}
 \end{figure}

\begin{figure} [h] 
  \includegraphics[width=0.5\textwidth] {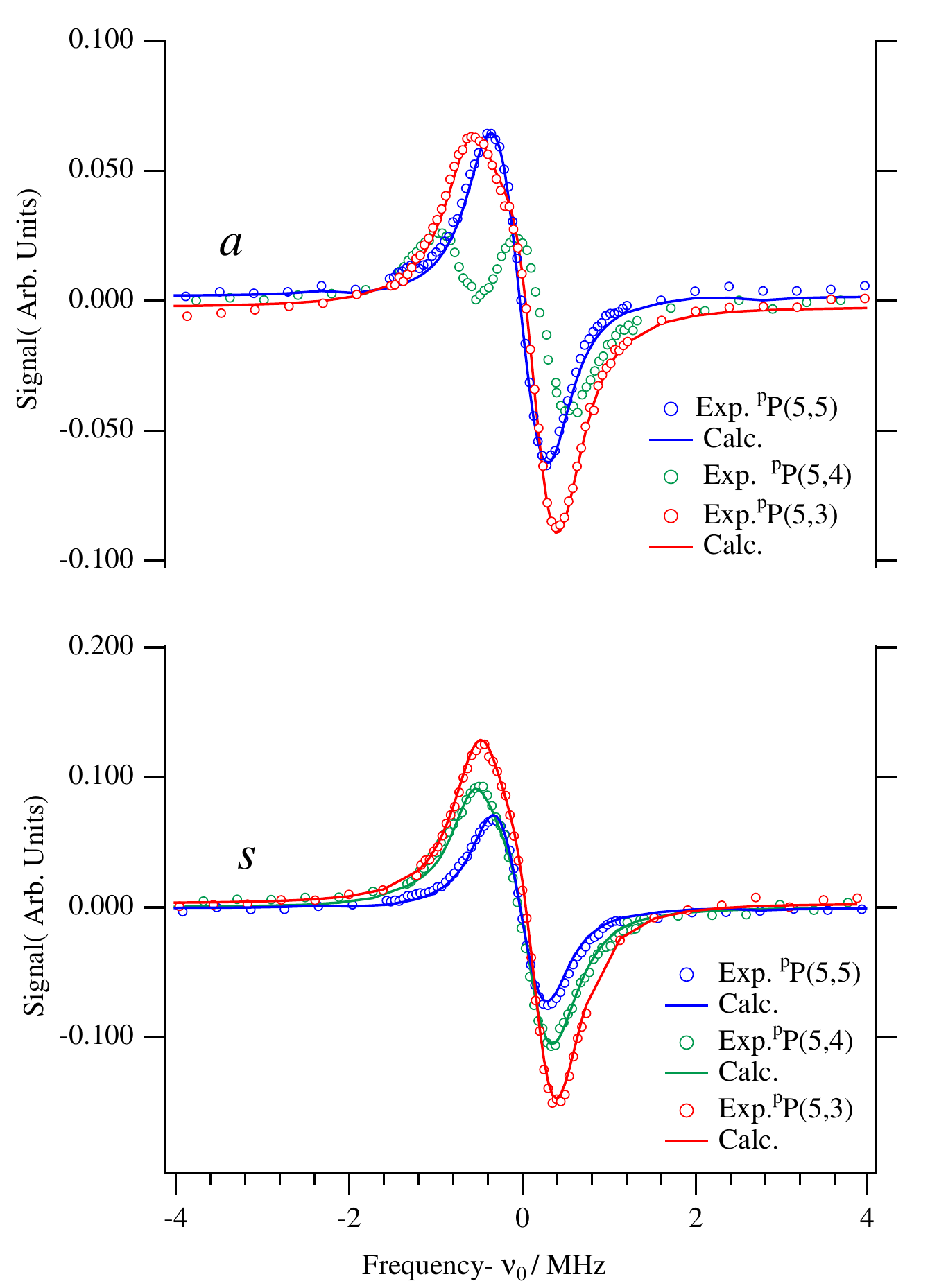} 
  \centering
 \caption{\footnotesize{Saturation dip line shapes for $P(5,3)$,  $P(5,4)$  and  $P(5,5$) transitions showing distorted or split line shapes due to hyperfine effects. The $P(5,4)_a$-component line is anomalous and deviates strongly from the line shape expected, based on the known quadrupolar hyperfine parameters. }}
 \label{fig4a}
 \end{figure}

   The observed $^pP(5,4)_a$ line shape shown in figure \ref{fig4a} is an outlier compared to the other $^pP(5,K)$ lines.  Comparison of the shape with simulations in figure \ref{fig3} suggests that the pattern more closely resembles that for a $^pQ$ or $^pR$ transition.  Alternatively, if the transition assignment is correct, the upper level quadrupole splitting has to be perturbed.   One other measured transition, $(v_1+2v_4)$, $^rP(7,5)_a$ was expected to show an extended splitting pattern such as the one in the lower right corner of figure \ref{fig3} but actually exhibits only a slightly asymmetrical shape, with a broader lower frequency lobe.  Finally, the feature at 6635.4971 \wn \,has been assigned\cite{Sung2012} to three different rotational transitions based on combination differences in the Doppler-limited spectra: $^rR(5,3)_a\, \mathrm{in\, the} (v_1+2v_4)$ band, $^rQ(6,4)_s\, \mathrm{in\, the} (v_1+v_3)$ band, and $^rQ(6,4)_s\, \mathrm{in\, the} (v_3+2v_4)$ band.  Searching near this frequency, we found only one sub-Doppler feature with a line shape that supports the first of these alternatives.  We attempted to record sub-Doppler spectra of other transitions connected to the upper state in all these transitions, but were unsuccessful, possibly due to insufficient sensitivity.  \\
 
   Absolute frequencies given in Table \ref{table2} are generally within the estimated errors of those reported by Sung \textit{et al.}\cite{Sung2012} and F\"{o}ldes \textit{et al.}\cite{Foldesi2014} when allowing for the fact that many of the new measurements are of components of overlapped features in the Doppler-limited spectra. The accuracy and precision of the present measurements places much tighter constraints on ground state combination differences in the ammonia spectrum in this region.  \\
 
   The present data permit the determination of six ground state combination differences. These are given in Table \ref{table3} where they are also compared to numbers derived from the published energy levels\cite{Urban1984} for the lowest inversion doublet levels.  Comparison shows that the present energy differences are systematically slightly larger than those derived from the published data, with the differences increasing with rotational energy.  Even so, the largest deviations are less than 1 MHz, and unresolved hyperfine splittings in the earlier data could be the major contributor to the differences.  Future improvements to the current spectrometer sensitivity will permit many more combination differences to be determined in these vibrational bands, and improvement in the accuracy of the rotational energy levels of ammonia could result from such measurements.  
 
\section{Conclusions}

The new data illustrate the precision that can be obtained from sub-Doppler measurements in the near-infrared.  The data provide accurate rest frequencies for transitions that are overlapped in the best Doppler-limited spectra, even at reduced temperatures, and we have illustrated how they can resolve ambiguities in spectral assignments.   The precision of lower state energy level combination differences derived from the measurements compares well with previous determinations derived from microwave and Doppler-limited far-infrared data. The observed quadrupole splittings in the sub-Doppler data provide a signature to help confirm rotational assignments because the quadrupole patterns are distinctive to a given rotational quantum number change.  All but a few of the 55+ sub-Doppler measurements exhibit observed quadrupole hyperfine patterns that match those expected based on predictions from the known splittings in the lowest inversion doublet levels.  While the upper, \Ep \,symmetry, level $(4,3)_a$ in the 1010(10) vibrational excited state accessed in the $^pP(5,4)_a$ transition appears perturbed, and lies close to the  $(5,4)_s$ level of 1002(02) with A$_2^{\prime}$ overall symmetry, these levels belong to different proton nuclear spin symmetries and, in ammonia, will be mixed by nuclear spin-rotation terms in the Hamiltonian. These are too small\cite{Cacciani2009} to account for the observed perturbation.  We therefore conclude that the rotational assignment is probably not correct in this case. \\   

In a similar vein, the assignment of the  $^rP(7,5)_a$ transition in the 0012(12) band at 6488.200 \wn \,is also probably incorrect because the observed hyperfine pattern does not match the expected.  Also, the upper level energy is calculated to be 6488.199520 + 463.70701 \wn \,from Table \ref{table1}.  This is 0.03 \wn \,from the value for the energy in the MARVEL database,\cite{AlDerzi2015} which is greater than expected based on the literature uncertainties.  There are no other, potentially perturbing, levels close to this energy in the database, so we conclude this assignment is also questionable. Finally, many of the data in Table \ref{table1} represent measurement of transitions that are overlapped in the Doppler-limited spectra.  For the most part, the sub-Doppler line measurements confirm the assignments in the literature.  But, for the line at 6635.4971 \wn, \, assigned to three separate rotation-vibration transtions\cite{Sung2012}, we could only find one sub-Doppler feature, that matched only one of the postulated assignments, raising questions about the other assignments. \\

  Efforts are underway to modify the spectrometer to increase its sensitivity and resolution still further. As it stands, sub-Doppler spectra can reliably be measured for unblended lines in ammonia with linestrengths, $S_{HITRAN}$, as low as 3$\times 10^{-22}$ cm.molecule$^{-1}$.  The major source of noise is due to vibrational and acoustic perturbations of the cavity and efforts are underway to counteract this.  The line widths could also be improved by increasing the beam waist diameter as described by Abe et al.\cite{Abe2014}  With these improvements, several of the assignment questions raised above could be definitively resolved.  In any case, the spectrum of ammonia in this region remains fertile ground for future sub-Doppler measurements.

\section{Acknowledgements}
Work at Brookhaven National Laboratory was carried out under Contract No. DE-SC0012704 with the U.S. Department of Energy, Office of Science, and supported by its Division of Chemical Sciences, Geosciences and Biosciences within the Office of Basic Energy Sciences.


\section{Figures and Tables}

\clearpage
\begingroup
\squeezetable
\begin{table}[h]
	\centering
	\caption{\footnotesize{Calculated quadrupole splittings for rotational levels of NH$_3$ in kHz}}
	\label{table1}
	\begin{tabular}{ccddd}
		\hline
		\hline
		\multicolumn{1}{c}{$J$} & \multicolumn{1}{c}{$K$} & \multicolumn{1}{c}{ $F_1=J-1$} & \multicolumn{1}{c}{ $F_1=J$} & \multicolumn{1}{c}{ $F_1=J+1$} \\ \hline    
		1  &   0  &     2043.8   &   -1021.9  &      204.4    \\   
		1  &   1  &    -1022.3   &     511.1  &     -102.2    \\   
		2  &   0  &     1020.8   &   -1020.8  &      291.7    \\   
		2  &   1  &      510.6   &    -510.6  &      145.9    \\   
		2  &   2  &    -1022.4   &    1022.4  &     -292.1    \\   
		3  &   0  &      815.4   &   -1019.3  &      339.8    \\   
		3  &   1  &      611.8   &    -764.7  &      254.9    \\   
		3  &   2  &       -0.0   &       0.0  &       -0.0    \\   
		3  &   3  &    -1022.7   &    1278.4  &     -426.1    \\   
		4  &   0  &      726.6   &   -1017.2  &      369.9    \\   
		4  &   1  &      617.8   &    -864.9  &      314.5     \\  
		4  &   2  &      291.1   &    -407.5  &      148.2     \\  
		4  &   3  &     -255.1   &     357.2  &     -129.9     \\  
		4  &   4  &    -1023.2   &    1432.5  &     -520.9     \\  
		5  &   0  &      676.4   &   -1014.6  &      390.2     \\  
		5  &   1  &      609.0   &    -913.4  &      351.3     \\  
		5  &   2  &      406.4   &    -609.6  &      234.5     \\  
		5  &   3  &       67.9   &    -101.8  &       39.2     \\  
		5  &   4  &     -408.3   &     612.4  &     -235.5     \\  
		5  &   5  &    -1024.0   &    1536.1  &     -590.8     \\  
		6  &   0  &      643.6   &   -1011.4  &      404.6     \\  
		6  &   1  &      597.9   &    -939.5  &      375.8     \\  
		6  &   2  &      460.4   &    -723.5  &      289.4     \\  
		6  &   3  &      230.6   &    -362.4  &      145.0     \\  
		6  &   4  &      -92.5   &     145.4  &      -58.1     \\  
		6  &   5  &     -510.5   &     802.1  &     -320.9     \\  
		6  &   6  &    -1025.1   &    1610.8  &     -644.3     \\  
		7  &   0  &      620.1   &   -1007.7  &      415.0     \\  
		7  &   1  &      587.1   &    -954.1  &      392.9     \\  
		7  &   2  &      488.0   &    -793.0  &      326.5     \\  
		7  &   3  &      322.2   &    -523.6  &      215.6     \\  
		7  &   4  &       89.1   &    -144.8  &       59.6     \\  
		7  &   5  &     -212.4   &     345.1  &     -142.1     \\  
		7  &   6  &     -583.7   &     948.5  &     -390.5     \\  
		7  &   7  &    -1026.3   &    1667.8  &     -686.7     \\  \hline \hline
	\end{tabular}
\end{table}
\endgroup

\clearpage
\begingroup
\squeezetable  
\begin{longtable*}{llllldddd}
	\caption{Frequency measured lines in the near-infrared spectrum of ammonia}\label{table2} \\
	
	\hline \hline \multicolumn{1}{c}{ \Np} & \multicolumn{1}{c}{ \Kp} & \multicolumn{1}{c}{ \Npp} & \multicolumn{1}{c}{ \Kpp} & \multicolumn{1}{l}{Vibration \supa}  & \multicolumn{1}{c} {\Epp \wn\,\supb} & \multicolumn{1}{c}{Frequency/GHz\,\supc} & \multicolumn{1}{c}{\wn} & \multicolumn{1}{c}{Previous/\wn\,\supd} \\  \hline
	\endfirsthead
	
	\multicolumn{9}{l}%
	{{\bfseries \tablename\ \thetable{} ...continued from previous page}} \\
	\hline \hline \multicolumn{1}{c}{ \Np} & \multicolumn{1}{c}{ \Kp} & \multicolumn{1}{c}{ \Npp} & \multicolumn{1}{c}{ \Kpp} & \multicolumn{1}{l}{Vibration \supa}  & \multicolumn{1}{c} {\Epp \wn\,\supb} & \multicolumn{1}{c}{Frequency/GHz\,\supc} & \multicolumn{1}{c}{\wn} & \multicolumn{1}{c}{Previous/\wn\,\supd} \\  \hline
	\endhead
	
	\multicolumn{9}{l} {{\textbf{...continued on next page}}} \\ \hline
	\endfoot
	
	\hline \hline
	\endlastfoot
	
	2  &  2  &  3 &  3   &  0012(12)s &  85.861 590  &   198 955.689 131  &   6 636.447 443 0  &	6636.448  \\
	2  &	2  &  3 &  3   &  0012(12)a &  85.657 810  &   198 963.757 520	&   6 636.716 575 4  &	6636.716  \\
	3  &  2  &  4	&  3   &  0012(12)a &  166.087 889  &  198 361.659 685  &   6 616.632 753 5  &  6616.633   \\   
	3  &  2  &  4	&  3   &  0012(12)s &  165.331 083  &  198 369.730 830  &   6 616.901 977 9  &  6616.902   \\   
	4  &  4  &  5	&  5   &  0012(12)a &  206.087 431  &  198 230.227 211  &   6 612.248 638 1  &  6612.249   \\   
	4  &  4  &  5	&  5   &  0012(12)s &  205.269 098  &  198 221.617 673  &   6 611.961 454 8  &  6611.962   \\   
	6  &  6  &  7	&  5   &  0012(12)a &  463.707 007  &  194 511.328 210$$^e$$ &   6 488.199 520 0  &  6488.200*  \\   
	3  &  2  &  4	&  3   &  1002(02)a &  166.087 888  &  195 073.628 560  &   6 506.955 840 8  &  6506.956   \\   
	3  &  2  &  4	&  3   &  1002(02)s &  165.331 083  &  195 113.933 570  &   6 508.300 271 2  &  6508.301   \\   
	4  &  2  &  5	&  3   &  1002(02)a &  265.226 620  &  194 549.591 940$$^e$$ &   6 489.475 860 7  &  6489.476   \\     
	4  &  2  &  5	&  3   &  1002(02)s &  264.516 615  &  194 558.668 340$$^e$$ &   6 489.778 616 8  &  6489.779   \\     
	4  &  4  &  5	&  5   &  1002(02)a &  206.087 431  &  195 165.624 892  &   6 510.024 508 1  &  6510.025   \\   
	5  &  4  &  6	&  5   &  1002(02)a &  325.127 182  &  194 607.591 710$$^e$$ &   6 491.410 524 7  &  6491.411   \\     
	6  &  4  &  5	&  3   &  1002(02)a &  265.226 620  &  198 927.198 079  &   6 635.497 083 8  &  6635.497   \\     
	1  &  0  &  1	&  1   &  1010(10)a &  16.963 349   &  198 243.691 814$$^e$$ &   6 612.697 768 9  &  6612.704*   \\    
	1  &  0  &  1	&  1   &  1010(10)s &  16.172 993   &  198 241.337 160$$^e$$ &   6 612.619 226 1  &  6612.619    \\    
	1  &  0  &  2	&  1   &  1010(10)s &  55.938 722   &  197 049.190 313$$^e$$ &   6 572.853 487 6  &  6572.854    \\    
	2  &  0  &  2	&  1   &  1010(10)a &  56.709 214   &  198 244.484 501$$^e$$ &   6 612.724 210 1  &  6612.726*   \\  
	2  &  0  &  2	&  1   &  1010(10)s &  55.938 722   &  198 244.855 520$$^e$$  &   6 612.736 585 9  &  6612.726*   \\  
	2  &  0  &  3	&  1   &  1010(10)a &  116.278 269  &  196 458.648 780  &   6 553.155 142 4  &  6553.155    \\  
	2  &  1  &  3	&  0   &  1010(10)s &  119.237 839  &  196 193.824 176$$^e$$ &   6 544.321 544 5  &  6544.322    \\  
	2  &	2  &  1 &  1   &  1010(10)s  & 16.172 933   &  198 945.712 286$$^e$$	&   6 636.114 651 2  &	6636.115 \\
	2  &	2  &  1 &  1   &  1010(10)a  & 16.172 933   &  198 948.000 247$$^e$$	&   6 636.190 969 4  &	6636.191 \\
	3  &  0  &  3	&  1   &  1010(10)a &  116.278 269  &  198 244.327 123  &   6 612.718 960 5  &  6612.726*   \\  
	3  &  0  &  3	&  1   &  1010(10)s &  115.536 605  &  198 247.689 375  &   6 612.831 113 1  &  6612.833*   \\  
	3  &  0  &  4	&  1   &  1010(10)a &  195.611 277  &  195 865.982 837  &   6 533.385 934 5  &  6533.386    \\  
	3  &  0  &  4	&  1   &  1010(10)s &  194.906 311  &  195 868.244 853  &   6 533.461 387 2  &  6533.461    \\  
	3  &  1  &  4	&  0   &  1010(10)a &  199.293 900  &  195 618.241 805  &   6 525.122 183 2  &  6525.122    \\  
	3  &  1  &  4	&  2   &  1010(10)a &  184.553 024  &  196 106.363 145  &   6 541.404 158 5  &  6541.405    \\  
	3  &  1  &  4	&  2   &  1010(10)s &  183.829 075  &  196 105.846 984  &   6 541.386 941 2  &  6541.387    \\  
	3  &  2  &  3	&  1   &  1010(10)a &  116.278 269  &  197 752.153 450  &   6 596.301 813 9  &  6596.302*   \\  
	3  &  2  &  4	&  1   &  1010(10)a &  195.611 277  &  195 373.809 131  &   6 516.968 786 8  &  6516.970    \\  
	3  &  2  &  4	&  1   &  1010(10)s &  194.906 311  &  195 375.305 456  &   6 517.018 698 9  &  6517.020    \\  
	3  &  2  &  4	&  3   &  1010(10)a &  166.087 888  &  196 328.038 380  &   6 548.798 448 4  &  6548.798    \\  
	3  &  2  &  4	&  3   &  1010(10)s &  165.331 083  &  196 322.408 681  &   6 548.610 661 9  &  6548.611    \\  
	4  &  0  &  4	&  1   &  1010(10)a &  195.611 277  &  198 242.993 623  &   6 612.674 479 7  &  6612.673*   \\  
	4  &  0  &  4	&  1   &  1010(10)s &  194.906 311  &  198 247.789 736  &   6 612.834 460 8  &  6612.824*   \\  
	4  &  0  &  5	&  1   &  1010(10)a &  294.629 992  &  195 274.486 618  &   6 513.655 744 4  &  6513.656    \\  
	4  &  0  &  5	&  1   &  1010(10)s &  293.968 256  &  195 277.986 831  &   6 513.772 498 9  &  6513.773    \\  
	4  &  1  &  5	&  2   &  1010(10)a &  283.616 663  &  195 531.648 422  &   6 522.233 738 8  &  6522.234    \\  
	4  &  1  &  5	&  2   &  1010(10)s &  282.937 143  &  195 523.772 061  &   6 521.971 011 7  &  6521.971    \\  
	4  &  2  &  5	&  3   &  1010(10)a &  265.226 620  &  195 727.557 554  &   6 528.768 564 1  &  6528.769    \\  
	4  &  2  &  5	&  3   &  1010(10)s &  264.516 615  &  195 740.182 541  &   6 529.189 688 3  &  6529.190    \\  
	4  &  3  &  5	&  4   &  1010(10)a &  239.408 225  &  195 994.734 505  &   6 537.680 627 9  &  6537.681    \\  
	4  &  3  &  5	&  4   &  1010(10)s &  238.652 596  &  195 962.122 314  &   6 536.592 802 3  &  6536.593    \\  
	4  &  4  &  5	&  5   &  1010(10)a &  206.087 431  &  196 136.937 270  &   6 542.424 001 5  &  6542.424    \\  
	4  &  4  &  5	&  5   &  1010(10)s &  205.269 098  &  196 142.933 714  &   6 542.624 021 4  &  6542.624    \\   
	5  &	1  &  5 &  2   &  1010)10)a &  283.616 663  &  198 372.131 726	&   6 616.982 063 2  &  6616.982    \\
	5  &  2  &  6	&  3   &  1010(10)a &  383.977 459  &  195 171.619 988  &   6 510.224 483 0  &  6510.225    \\  
	5  &  4  &  6	&  5   &  1010(10)a &  325.127 182  &  195 560.864 922  &   6 523.208 296 4  &  6523.208    \\  
	5  &  4  &  6	&  5   &  1010(10)s &  324.368 904  &  195 583.055 630  &   6 523.948 498 7  &  6523.949*   \\  
	5  &  5  &  6	&  6   &  1010(10)a &  284.410 125  &  195 727.661 582  &   6 528.772 034 1  &  6528.772    \\  
	5  &  5  &  6	&  6   &  1010(10)s &  283.574 345  &  195 731.505 651  &   6 528.900 258 4  &  6528.901    \\  
	6  &  0  &  6	&  1   &  1010(10)s &  412.624 301  &  198 242.568 537  &   6 612.660 300 4  &  6612.653*   \\  
	6  &  2  &  7	&  3   &  1010(10)a &  522.222 914  &  194 500.699 730  &   6 487.844 992 1  &  6487.845    \\  
	6  &  2  &  7	&  3   &  1010(10)s &  521.621 923  &  194 570.952 480  &   6 490.188 371 6  &  6490.189    \\  
	6  &  5  &  7	&  6   &  1010(10)s &  422.458 103  &  195 194.649 722  &   6 510.992 672 2  &  6510.993    \\  
	7  &  0  &  7	&  1   &  1010(10)s &  550.758 585  &  198 257.822 506  &   6 613.169 118 0  &  6613.169    \\  
	
\end{longtable*}
\endgroup

\footnotesize   
{\parindent0pt   
	Footnotes: \\ 
	\textit{a}. Labels are the upper and lower level rotational quantum numbers,  $v_1 v_2 v_3 v_4(|l_3|,|l_4|)$ for the upper level in the transition and inversion symmetry (a or s) of the levels involved. \\ 
	\textit{b}. Lower state energy from Urban \textit{et al.}\cite{Urban1984}  \\ 
	\textit{c}. Measured frequencies have an absolute error estimated to be less than 20 kHz unless indicated.  See text for details. The corresponding wavenumbers are also given in the following column. \\
	\textit{d}. Previous measurements from Sung \textit{et al.}\cite{Sung2012} and F\"{o}ldes \textit{et al.} \cite{Foldesi2014} and  references therein. Transitions marked with an asterisk are blended in the Doppler-limited spectrum.\\ 
}  
\textit{e}. Line center measurement less accurate due to poor signal-to-noise or complex hyperfine pattern.  Estimated errors up to $\pm$30kHz.


\begingroup
\squeezetable
\begin{table}[h]
	\centering
	\caption{\footnotesize{Ground state combination differences determined from the measurements}}
	\label{table3}
	\begin{tabular}{cdd}
		\hline
		\hline
		\multicolumn{1}{c}{($J,K$) - ($J,K$)} & \multicolumn{1}{c}{Frequency/GHz} & \multicolumn{1}{c}{Urban \textit{et al.}\cite{Urban1984}} \\ \hline 
		(2,1) - (1,1)s &   1 192.146 847 & 1 192.146 564  \\
		(3,1) - (2,1)a &   1 785.835 721 & 1 785.835 342  \\
		(4,1) - (3,1)s &  2 379.444 522  & 2 379.443 925  \\ 
		(4,1) - (3,1)a &  2 378.344 319  & 2 378.343 747  \\ 
		(5,1) - (4,1)a &  2 968.507 005  & 2 968.506 396  \\ 
		(5,1) - (4,1)s &  2 969.802 905	 & 2 969.802 399  \\ \hline \hline
	\end{tabular}
\end{table}
\endgroup


\begin{thebibliography}{24}%
\makeatletter
\providecommand \@ifxundefined [1]{%
 \@ifx{#1\undefined}
}%
\providecommand \@ifnum [1]{%
 \ifnum #1\expandafter \@firstoftwo
 \else \expandafter \@secondoftwo
 \fi
}%
\providecommand \@ifx [1]{%
 \ifx #1\expandafter \@firstoftwo
 \else \expandafter \@secondoftwo
 \fi
}%
\providecommand \natexlab [1]{#1}%
\providecommand \enquote  [1]{``#1''}%
\providecommand \bibnamefont  [1]{#1}%
\providecommand \bibfnamefont [1]{#1}%
\providecommand \citenamefont [1]{#1}%
\providecommand \href@noop [0]{\@secondoftwo}%
\providecommand \href [0]{\begingroup \@sanitize@url \@href}%
\providecommand \@href[1]{\@@startlink{#1}\@@href}%
\providecommand \@@href[1]{\endgroup#1\@@endlink}%
\providecommand \@sanitize@url [0]{\catcode `\\12\catcode `\$12\catcode
  `\&12\catcode `\#12\catcode `\^12\catcode `\_12\catcode `\%12\relax}%
\providecommand \@@startlink[1]{}%
\providecommand \@@endlink[0]{}%
\providecommand \url  [0]{\begingroup\@sanitize@url \@url }%
\providecommand \@url [1]{\endgroup\@href {#1}{\urlprefix }}%
\providecommand \urlprefix  [0]{URL }%
\providecommand \Eprint [0]{\href }%
\providecommand \doibase [0]{http://dx.doi.org/}%
\providecommand \selectlanguage [0]{\@gobble}%
\providecommand \bibinfo  [0]{\@secondoftwo}%
\providecommand \bibfield  [0]{\@secondoftwo}%
\providecommand \translation [1]{[#1]}%
\providecommand \BibitemOpen [0]{}%
\providecommand \bibitemStop [0]{}%
\providecommand \bibitemNoStop [0]{.\EOS\space}%
\providecommand \EOS [0]{\spacefactor3000\relax}%
\providecommand \BibitemShut  [1]{\csname bibitem#1\endcsname}%
\let\auto@bib@innerbib\@empty
\bibitem [{\citenamefont {Kroto}(1992)}]{Kroto}%
  \BibitemOpen
  \bibfield  {author} {\bibinfo {author} {\bibfnamefont {H.~W.}\ \bibnamefont
  {Kroto}},\ }\href@noop {} {\emph {\bibinfo {title} {Molecular Rotation
  Spectra}}}\ (\bibinfo  {publisher} {Dover Publications, NY},\ \bibinfo {year}
  {1992})\BibitemShut {NoStop}%
\bibitem [{\citenamefont {Townes}\ and\ \citenamefont
  {Schawlow}(1975)}]{Townes}%
  \BibitemOpen
  \bibfield  {author} {\bibinfo {author} {\bibfnamefont {C.~H.}\ \bibnamefont
  {Townes}}\ and\ \bibinfo {author} {\bibfnamefont {A.~L.}\ \bibnamefont
  {Schawlow}},\ }\href@noop {} {\emph {\bibinfo {title} {{Microwave
  Spectroscopy}}}}\ (\bibinfo  {publisher} {Dover Publications, NY},\ \bibinfo
  {year} {1975})\BibitemShut {NoStop}%
\bibitem [{\citenamefont {Bunker}\ and\ \citenamefont
  {Jensen}(1998)}]{Bunkerbook}%
  \BibitemOpen
  \bibfield  {author} {\bibinfo {author} {\bibfnamefont {P.~R.}\ \bibnamefont
  {Bunker}}\ and\ \bibinfo {author} {\bibfnamefont {P.}~\bibnamefont
  {Jensen}},\ }\href@noop {} {\emph {\bibinfo {title} {{Molecular Symmetry and
  Spectroscopy, Second Edition}}}}\ (\bibinfo  {publisher} {NRC Press,
  Ottawa},\ \bibinfo {year} {1998})\BibitemShut {NoStop}%
\bibitem [{\citenamefont {Sung}\ \emph {et~al.}(2012)\citenamefont {Sung},
  \citenamefont {Brown}, \citenamefont {Huang}, \citenamefont {Schwenke},
  \citenamefont {Lee}, \citenamefont {Coy},\ and\ \citenamefont
  {Lehmann}}]{Sung2012}%
  \BibitemOpen
  \bibfield  {author} {\bibinfo {author} {\bibfnamefont {K.}~\bibnamefont
  {Sung}}, \bibinfo {author} {\bibfnamefont {L.~R.}\ \bibnamefont {Brown}},
  \bibinfo {author} {\bibfnamefont {X.}~\bibnamefont {Huang}}, \bibinfo
  {author} {\bibfnamefont {D.~W.}\ \bibnamefont {Schwenke}}, \bibinfo {author}
  {\bibfnamefont {T.~J.}\ \bibnamefont {Lee}}, \bibinfo {author} {\bibfnamefont
  {S.~L.}\ \bibnamefont {Coy}}, \ and\ \bibinfo {author} {\bibfnamefont
  {K.~K.}\ \bibnamefont {Lehmann}},\ }\href {\doibase
  http://dx.doi.org/10.1016/j.jqsrt.2012.02.037} {\bibfield  {journal}
  {\bibinfo  {journal} {J. Quant. Spectr. and Rad.Transf}\ }\textbf {\bibinfo
  {volume} {113(11)}},\ \bibinfo {pages} {1066 } (\bibinfo {year}
  {2012})}\BibitemShut {NoStop}%
\bibitem [{\citenamefont {Cacciani}\ \emph {et~al.}(2012)\citenamefont
  {Cacciani}, \citenamefont {Cermak}, \citenamefont {Cosl\'eou}, \citenamefont
  {Khelkhal}, \citenamefont {Jeseck},\ and\ \citenamefont
  {Michaut}}]{Cacciani2012}%
  \BibitemOpen
  \bibfield  {author} {\bibinfo {author} {\bibfnamefont {P.}~\bibnamefont
  {Cacciani}}, \bibinfo {author} {\bibfnamefont {P.}~\bibnamefont {Cermak}},
  \bibinfo {author} {\bibfnamefont {J.}~\bibnamefont {Cosl\'eou}}, \bibinfo
  {author} {\bibfnamefont {M.}~\bibnamefont {Khelkhal}}, \bibinfo {author}
  {\bibfnamefont {P.}~\bibnamefont {Jeseck}}, \ and\ \bibinfo {author}
  {\bibfnamefont {X.}~\bibnamefont {Michaut}},\ }\href {\doibase
  http://dx.doi.org/10.1016/j.jqsrt.2012.02.026} {\bibfield  {journal}
  {\bibinfo  {journal} {J. Quant. Spectr. and Rad.Transf}\ }\textbf {\bibinfo
  {volume} {113}},\ \bibinfo {pages} {1084 } (\bibinfo {year}
  {2012})}\BibitemShut {NoStop}%
\bibitem [{\citenamefont {Foldes}\ \emph {et~al.}(2014)\citenamefont {Foldes},
  \citenamefont {Golebiowski}, \citenamefont {Herman}, \citenamefont {Softley},
  \citenamefont {DiLonardo},\ and\ \citenamefont {Fusina}}]{Foldesi2014}%
  \BibitemOpen
  \bibfield  {author} {\bibinfo {author} {\bibfnamefont {T.}~\bibnamefont
  {Foldes}}, \bibinfo {author} {\bibfnamefont {D.}~\bibnamefont {Golebiowski}},
  \bibinfo {author} {\bibfnamefont {M.}~\bibnamefont {Herman}}, \bibinfo
  {author} {\bibfnamefont {T.~P.}\ \bibnamefont {Softley}}, \bibinfo {author}
  {\bibfnamefont {G.}~\bibnamefont {DiLonardo}}, \ and\ \bibinfo {author}
  {\bibfnamefont {L.}~\bibnamefont {Fusina}},\ }\href@noop {} {\bibfield
  {journal} {\bibinfo  {journal} {Molec. Phys.}\ }\textbf {\bibinfo {volume}
  {112}},\ \bibinfo {pages} {2407} (\bibinfo {year} {2014})}\BibitemShut
  {NoStop}%
\bibitem [{\citenamefont {AlDerzi}\ \emph {et~al.}(2015)\citenamefont
  {AlDerzi}, \citenamefont {Furtenbacher}, \citenamefont {Tennyson},
  \citenamefont {Yurchenko},\ and\ \citenamefont {Cs\'asz\'ar}}]{AlDerzi2015}%
  \BibitemOpen
  \bibfield  {author} {\bibinfo {author} {\bibfnamefont {A.~R.}\ \bibnamefont
  {AlDerzi}}, \bibinfo {author} {\bibfnamefont {T.}~\bibnamefont
  {Furtenbacher}}, \bibinfo {author} {\bibfnamefont {J.}~\bibnamefont
  {Tennyson}}, \bibinfo {author} {\bibfnamefont {S.~N.}\ \bibnamefont
  {Yurchenko}}, \ and\ \bibinfo {author} {\bibfnamefont {A.~G.}\ \bibnamefont
  {Cs\'asz\'ar}},\ }\href {\doibase
  http://dx.doi.org/10.1016/j.jqsrt.2015.03.034} {\bibfield  {journal}
  {\bibinfo  {journal} {J. Quant. Spectr. and Rad.Transf}\ }\textbf {\bibinfo
  {volume} {161}},\ \bibinfo {pages} {117 } (\bibinfo {year}
  {2015})}\BibitemShut {NoStop}%
\bibitem [{\citenamefont {Lehmann}\ and\ \citenamefont {Coy}(1988)}]{Lehmann}%
  \BibitemOpen
  \bibfield  {author} {\bibinfo {author} {\bibfnamefont {K.~K.}\ \bibnamefont
  {Lehmann}}\ and\ \bibinfo {author} {\bibfnamefont {S.~L.}\ \bibnamefont
  {Coy}},\ }\href@noop {} {\bibfield  {journal} {\bibinfo  {journal} {J. Chem.
  Soc., Faraday Trans. 2}\ }\textbf {\bibinfo {volume} {84(9)}},\ \bibinfo
  {pages} {1389} (\bibinfo {year} {1988})}\BibitemShut {NoStop}%
\bibitem [{\citenamefont {Douglas}\ and\ \citenamefont
  {Hollas}(1961)}]{Douglas}%
  \BibitemOpen
  \bibfield  {author} {\bibinfo {author} {\bibfnamefont {A.~E.}\ \bibnamefont
  {Douglas}}\ and\ \bibinfo {author} {\bibfnamefont {J.~M.}\ \bibnamefont
  {Hollas}},\ }\href@noop {} {\bibfield  {journal} {\bibinfo  {journal} {Can.
  J. Phys.}\ }\textbf {\bibinfo {volume} {39(4)}},\ \bibinfo {pages} {479}
  (\bibinfo {year} {1961})}\BibitemShut {NoStop}%
\bibitem [{\citenamefont {Czajkowski}\ \emph {et~al.}(2009)\citenamefont
  {Czajkowski}, \citenamefont {Alcock}, \citenamefont {Bernard}, \citenamefont
  {Madej}, \citenamefont {Corrigan},\ and\ \citenamefont
  {Cheurov}}]{Czajkowski2009}%
  \BibitemOpen
  \bibfield  {author} {\bibinfo {author} {\bibfnamefont {A.}~\bibnamefont
  {Czajkowski}}, \bibinfo {author} {\bibfnamefont {A.~J.}\ \bibnamefont
  {Alcock}}, \bibinfo {author} {\bibfnamefont {J.~E.}\ \bibnamefont {Bernard}},
  \bibinfo {author} {\bibfnamefont {A.~A.}\ \bibnamefont {Madej}}, \bibinfo
  {author} {\bibfnamefont {M.}~\bibnamefont {Corrigan}}, \ and\ \bibinfo
  {author} {\bibfnamefont {S.}~\bibnamefont {Cheurov}},\ }\href@noop {}
  {\bibfield  {journal} {\bibinfo  {journal} {Opt. Exp}\ }\textbf {\bibinfo
  {volume} {17}},\ \bibinfo {pages} {9258} (\bibinfo {year}
  {2009})}\BibitemShut {NoStop}%
\bibitem [{\citenamefont {Berden}, \citenamefont {Peeters},\ and\ \citenamefont
  {Meijer}(1999)}]{Berden1999}%
  \BibitemOpen
  \bibfield  {author} {\bibinfo {author} {\bibfnamefont {G.}~\bibnamefont
  {Berden}}, \bibinfo {author} {\bibfnamefont {R.}~\bibnamefont {Peeters}}, \
  and\ \bibinfo {author} {\bibfnamefont {G.}~\bibnamefont {Meijer}},\
  }\href@noop {} {\bibfield  {journal} {\bibinfo  {journal} {Chem. Phys.
  Letts.}\ }\textbf {\bibinfo {volume} {307}},\ \bibinfo {pages} {131}
  (\bibinfo {year} {1999})}\BibitemShut {NoStop}%
\bibitem [{\citenamefont {Lees}, \citenamefont {Li},\ and\ \citenamefont
  {Xu}(2008)}]{Lees2008}%
  \BibitemOpen
  \bibfield  {author} {\bibinfo {author} {\bibfnamefont {R.}~\bibnamefont
  {Lees}}, \bibinfo {author} {\bibfnamefont {L.}~\bibnamefont {Li}}, \ and\
  \bibinfo {author} {\bibfnamefont {L.-H.}\ \bibnamefont {Xu}},\ }\href@noop {}
  {\bibfield  {journal} {\bibinfo  {journal} {J. Mol. Spectrosc}\ }\textbf
  {\bibinfo {volume} {251}},\ \bibinfo {pages} {241 } (\bibinfo {year}
  {2008})}\BibitemShut {NoStop}%
\bibitem [{\citenamefont {Kukolich}(1967{\natexlab{a}})}]{Kukolich67}%
  \BibitemOpen
  \bibfield  {author} {\bibinfo {author} {\bibfnamefont {S.~G.}\ \bibnamefont
  {Kukolich}},\ }\href@noop {} {\bibfield  {journal} {\bibinfo  {journal}
  {Phys. Rev.}\ }\textbf {\bibinfo {volume} {156}},\ \bibinfo {pages} {83}
  (\bibinfo {year} {1967}{\natexlab{a}})}\BibitemShut {NoStop}%
\bibitem [{\citenamefont {Kukolich}(1967{\natexlab{b}})}]{Kukolich68}%
  \BibitemOpen
  \bibfield  {author} {\bibinfo {author} {\bibfnamefont {S.~G.}\ \bibnamefont
  {Kukolich}},\ }\href@noop {} {\bibfield  {journal} {\bibinfo  {journal}
  {Phys. Rev.}\ }\textbf {\bibinfo {volume} {172}},\ \bibinfo {pages} {59}
  (\bibinfo {year} {1967}{\natexlab{b}})}\BibitemShut {NoStop}%
\bibitem [{\citenamefont {Kukolich}\ and\ \citenamefont
  {Wofsy}(1970)}]{Kukolich70}%
  \BibitemOpen
  \bibfield  {author} {\bibinfo {author} {\bibfnamefont {S.~G.}\ \bibnamefont
  {Kukolich}}\ and\ \bibinfo {author} {\bibfnamefont {S.~C.}\ \bibnamefont
  {Wofsy}},\ }\href@noop {} {\bibfield  {journal} {\bibinfo  {journal} {J.
  Chem. Phys.}\ }\textbf {\bibinfo {volume} {52}},\ \bibinfo {pages} {5477}
  (\bibinfo {year} {1970})}\BibitemShut {NoStop}%
\bibitem [{\citenamefont {Hougen}(1972)}]{Hougen72}%
  \BibitemOpen
  \bibfield  {author} {\bibinfo {author} {\bibfnamefont {J.~T.}\ \bibnamefont
  {Hougen}},\ }\href@noop {} {\bibfield  {journal} {\bibinfo  {journal} {J.
  Chem. Phys.}\ }\textbf {\bibinfo {volume} {57}},\ \bibinfo {pages} {4207}
  (\bibinfo {year} {1972})}\BibitemShut {NoStop}%
\bibitem [{\citenamefont {Cubillas}, \citenamefont {Hald},\ and\ \citenamefont
  {Petersen}(2008)}]{Cubillas2008}%
  \BibitemOpen
  \bibfield  {author} {\bibinfo {author} {\bibfnamefont {A.~M.}\ \bibnamefont
  {Cubillas}}, \bibinfo {author} {\bibfnamefont {J.}~\bibnamefont {Hald}}, \
  and\ \bibinfo {author} {\bibfnamefont {J.~C.}\ \bibnamefont {Petersen}},\
  }\href@noop {} {\bibfield  {journal} {\bibinfo  {journal} {Opt. Expr..}\
  }\textbf {\bibinfo {volume} {16}},\ \bibinfo {pages} {3976} (\bibinfo {year}
  {2008})}\BibitemShut {NoStop}%
\bibitem [{\citenamefont {Petersen}\ and\ \citenamefont
  {Hald}(2010)}]{Petersen2010}%
  \BibitemOpen
  \bibfield  {author} {\bibinfo {author} {\bibfnamefont {J.~C.}\ \bibnamefont
  {Petersen}}\ and\ \bibinfo {author} {\bibfnamefont {J.}~\bibnamefont
  {Hald}},\ }\href@noop {} {\bibfield  {journal} {\bibinfo  {journal} {Opt.
  Expr..}\ }\textbf {\bibinfo {volume} {18}},\ \bibinfo {pages} {7955}
  (\bibinfo {year} {2010})}\BibitemShut {NoStop}%
\bibitem [{\citenamefont {Cacciani}\ \emph {et~al.}(2009)\citenamefont
  {Cacciani}, \citenamefont {Cosl\'eou}, \citenamefont {Khelkhal},
  \citenamefont {Tudorie}, \citenamefont {Puzzurini},\ and\ \citenamefont
  {Pracna}}]{Cacciani2009}%
  \BibitemOpen
  \bibfield  {author} {\bibinfo {author} {\bibfnamefont {P.}~\bibnamefont
  {Cacciani}}, \bibinfo {author} {\bibfnamefont {J.}~\bibnamefont {Cosl\'eou}},
  \bibinfo {author} {\bibfnamefont {M.}~\bibnamefont {Khelkhal}}, \bibinfo
  {author} {\bibfnamefont {M.}~\bibnamefont {Tudorie}}, \bibinfo {author}
  {\bibfnamefont {C.}~\bibnamefont {Puzzurini}}, \ and\ \bibinfo {author}
  {\bibfnamefont {P.}~\bibnamefont {Pracna}},\ }\href@noop {} {\bibfield
  {journal} {\bibinfo  {journal} {Phys. Rev. A}\ }\textbf {\bibinfo {volume}
  {80}},\ \bibinfo {pages} {042507(10)} (\bibinfo {year} {2009})}\BibitemShut
  {NoStop}%
\bibitem [{\citenamefont {Twagirayezu}\ \emph {et~al.}(2015)\citenamefont
  {Twagirayezu}, \citenamefont {Cich}, \citenamefont {Sears}, \citenamefont
  {McRaven},\ and\ \citenamefont {Hall}}]{Twagirayezu1}%
  \BibitemOpen
  \bibfield  {author} {\bibinfo {author} {\bibfnamefont {S.}~\bibnamefont
  {Twagirayezu}}, \bibinfo {author} {\bibfnamefont {M.~J.}\ \bibnamefont
  {Cich}}, \bibinfo {author} {\bibfnamefont {T.~J.}\ \bibnamefont {Sears}},
  \bibinfo {author} {\bibfnamefont {C.~P.}\ \bibnamefont {McRaven}}, \ and\
  \bibinfo {author} {\bibfnamefont {G.~E.}\ \bibnamefont {Hall}},\ }\href@noop
  {} {\bibfield  {journal} {\bibinfo  {journal} {J. Molec. Spectrosc.}\
  }\textbf {\bibinfo {volume} {316}},\ \bibinfo {pages} {64} (\bibinfo {year}
  {2015})}\BibitemShut {NoStop}%
\bibitem [{\citenamefont {Axner}, \citenamefont {Kluczynski},\ and\
  \citenamefont {Lindberg}(2001)}]{Axner2001}%
  \BibitemOpen
  \bibfield  {author} {\bibinfo {author} {\bibfnamefont {O.}~\bibnamefont
  {Axner}}, \bibinfo {author} {\bibfnamefont {P.}~\bibnamefont {Kluczynski}}, \
  and\ \bibinfo {author} {\bibfnamefont {{\"O}.~M.}\ \bibnamefont {Lindberg}},\
  }\href@noop {} {\bibfield  {journal} {\bibinfo  {journal} {J. Quant. Spectr.
  and Rad.Transf.}\ }\textbf {\bibinfo {volume} {68}},\ \bibinfo {pages} {299 }
  (\bibinfo {year} {2001})}\BibitemShut {NoStop}%
\bibitem [{\citenamefont {Demtr{\"o}der}(1996)}]{Demtroder}%
  \BibitemOpen
  \bibfield  {author} {\bibinfo {author} {\bibfnamefont {W.}~\bibnamefont
  {Demtr{\"o}der}},\ }\href@noop {} {\emph {\bibinfo {title} {Laser
  Spectroscopy: Basic Concepts and Instrumentation}}}\ (\bibinfo  {publisher}
  {Springer-Verlag},\ \bibinfo {year} {1996})\BibitemShut {NoStop}%
\bibitem [{\citenamefont {Urban}\ \emph {et~al.}(1984)\citenamefont {Urban},
  \citenamefont {D'Cuhna}, \citenamefont {Rao},\ and\ \citenamefont {Papou{\u
  s}ek}}]{Urban1984}%
  \BibitemOpen
  \bibfield  {author} {\bibinfo {author} {\bibfnamefont {{\u S}.}~\bibnamefont
  {Urban}}, \bibinfo {author} {\bibfnamefont {R.}~\bibnamefont {D'Cuhna}},
  \bibinfo {author} {\bibfnamefont {K.~N.}\ \bibnamefont {Rao}}, \ and\
  \bibinfo {author} {\bibfnamefont {D.}~\bibnamefont {Papou{\u s}ek}},\
  }\href@noop {} {\bibfield  {journal} {\bibinfo  {journal} {Can. J. Phys.}\
  }\textbf {\bibinfo {volume} {62}},\ \bibinfo {pages} {1775} (\bibinfo {year}
  {1984})}\BibitemShut {NoStop}%
\bibitem [{\citenamefont {Abe}\ \emph {et~al.}(2014)\citenamefont {Abe},
  \citenamefont {Iwakuni}, \citenamefont {Okubo},\ and\ \citenamefont
  {Sasada}}]{Abe2014}%
  \BibitemOpen
  \bibfield  {author} {\bibinfo {author} {\bibfnamefont {M.}~\bibnamefont
  {Abe}}, \bibinfo {author} {\bibfnamefont {K.}~\bibnamefont {Iwakuni}},
  \bibinfo {author} {\bibfnamefont {S.}~\bibnamefont {Okubo}}, \ and\ \bibinfo
  {author} {\bibfnamefont {H.}~\bibnamefont {Sasada}},\ }\href@noop {}
  {\bibfield  {journal} {\bibinfo  {journal} {Opt. Letts.}\ }\textbf {\bibinfo
  {volume} {39}},\ \bibinfo {pages} {5277} (\bibinfo {year}
  {2014})}\BibitemShut {NoStop}%
\end{thebibliography}
%

\end{document}